\documentclass[11pt]{article}

\usepackage[letterpaper,margin=1in]{geometry}
\usepackage{amsmath, amssymb, mathtools}
\usepackage{graphicx}
\usepackage{float}
\usepackage{subcaption}
\usepackage{booktabs}
\usepackage[hidelinks]{hyperref}
\usepackage[capitalise]{cleveref}
\usepackage{authblk}
\usepackage{array}
\usepackage{makecell}
\usepackage[table]{xcolor}
\usepackage{threeparttable}
\usepackage{tabularx}
\usepackage{multirow}
\usepackage{setspace}

\definecolor{headerblue}{HTML}{2F5FA8}
\definecolor{lightblue}{HTML}{EAF1FB}
\definecolor{ioccshade}{HTML}{FFF2CC}
\definecolor{ioccorange}{HTML}{D95F02}
\definecolor{passblue}{HTML}{DDEEFF}
\newcolumntype{C}[1]{>{\centering\arraybackslash}m{#1}}
\newcolumntype{L}[1]{>{\raggedright\arraybackslash}m{#1}}
\newcolumntype{G}[1]{>{\columncolor{ioccshade}[\tabcolsep][\tabcolsep]\centering\arraybackslash}m{#1}}

\newcommand{\vect}[1]{\boldsymbol{#1}}
\newcommand{\matr}[1]{\boldsymbol{#1}}
\newcommand{\tr}{\mathrm{tr}}

\title{Tulip-Shaped Orbits for Lunar South-Pole PNT \\ and Direct-to-Earth Relay Missions}

\author[]{Darin C.\ Koblick\thanks{Corresponding author: \texttt{Darin@coorbital.com}} }
\author[]{Michael Casey}
\affil[]{Coorbital, Inc.}

\date{July 2026}

\begin{document}
\maketitle

\begin{abstract}
\noindent This geometric study evaluates a compact seven-petal, \(6\!:\!5\)-resonant tulip-shaped orbit constellation for lunar south-pole positioning, navigation, and timing (PNT) and direct-to-Earth relay services. The tulip-shaped orbits are compared against elliptical lunar frozen orbit (ELFO) constellations over the NASA LunaNet Service Volume II (SV2), covering lunar latitudes south of $-75^\circ$. We compare a six-satellite tulip baseline with a minimum-cost five-satellite variant; both use the same shared three-body orbit and differ only in satellite count and along-track phasing. Performance is scored against three Initial Operating Capability~C (IOC-C) metrics: line-of-sight (LOS) link availability, Lunar Augmented Navigation System (LANS) geometric dilution of precision (GDOP) at $\mathrm{GDOP}<6$, and daily extravehicular activity (EVA) usable-PNT windows. Both tulip constellations satisfy all three IOC-C metrics across SV2. The six-satellite configuration meets requirements with wide margin: 75\% worst-point daily $\mathrm{GDOP}<6$ availability and 18~h of daily EVA support. The five-satellite variant also passes, but with thinner margin: 44\% availability and 10~h of EVA support. Unlike the ELFO configurations, each spacecraft in this tulip-shaped orbit configuration maintains continuous Earth line of sight, providing continuous geometric opportunity for direct single-hop Earth relay. This persistent Earth visibility is paired with a three-body orbit, reducing $\Delta V$ requirements for initial phasing and reconstitution maneuvers.
\end{abstract}

%==============================================================================
\section{Introduction}
%==============================================================================
NASA's Artemis program, commercial lunar missions, and international lunar initiatives create immediate demand for lunar surface positioning, navigation, and timing (PNT) and communications relay services. South-polar users see Earth only a few degrees above the horizon~\cite{li2024design}, so reliable navigation and data return require orbiting infrastructure~\cite{nasa2022lunanet,nasa2023lcrns,esa2024moonlight}. NASA's LunaNet / Lunar Communications Relay and Navigation Systems (LCRNS) effort addresses this need through phased relay, PNT, and constellation-geometry requirements~\cite{lcrns2022srd}.

This paper evaluates constellations of \emph{tulip-shaped orbits} against the Initial Operating Capability~C (IOC-C) requirements because IOC-C is the first LunaNet increment that strongly drives constellation geometry. At IOC-C, the network must provide Service Volume~II (SV2) coverage, multiple simultaneous relay links, and four geometrically diverse Augmented Forward Signal (AFS) / Lunar Augmented Navigation System (LANS) links with GDOP below~6~\cite{lcrns2022srd}. \Cref{tab:lcrns_sizing} summarizes the SRD constellation-sizing requirements and highlights the IOC-C benchmark used in this study.

\begin{table}[H]
\centering
\caption{Constellation sizing requirements from the NASA LCRNS SRD, with the IOC-C increment highlighted. ``Min coverage'' is the minimum coverage of the corresponding service volume specified by the SRD~\cite{lcrns2022srd}.}
\label{tab:lcrns_sizing}
\small
\renewcommand{\arraystretch}{1.18}
\setlength{\tabcolsep}{3pt}

\begin{threeparttable}
\resizebox{\linewidth}{!}{%
\begin{tabular}{
|C{3.2cm} |
C{1.10cm} C{1.10cm} C{1.15cm} |
C{1.10cm} C{1.10cm} C{1.10cm} C{1.0cm} |
G{1.10cm} G{1.10cm} G{2.50cm} |
C{1.10cm} C{1.10cm} C{2.00cm} |
}

\multicolumn{1}{|>{\columncolor{headerblue}}c|}{\textcolor{white}{\textbf{Service Type}}} &
\multicolumn{3}{>{\columncolor{headerblue}}c|}{\textcolor{white}{\textbf{IOC-A}}} &
\multicolumn{4}{>{\columncolor{headerblue}}c|}{\textcolor{white}{\textbf{IOC-B}}} &
\multicolumn{3}{>{\columncolor{ioccorange}}c|}{\textcolor{white}{\textbf{IOC-C}}} &
\multicolumn{3}{>{\columncolor{headerblue}}c|}{\textcolor{white}{\textbf{EOC}}} \\ \hline

\cellcolor{lightblue}\textbf{} &
\cellcolor{lightblue}\textbf{Ka} &
\cellcolor{lightblue}\textbf{S} &
\cellcolor{lightblue}\textbf{AFS} &
\cellcolor{lightblue}\textbf{Ka} &
\cellcolor{lightblue}\textbf{S} &
\multicolumn{2}{>{\columncolor{lightblue}}c|}{\textbf{AFS}} &
\cellcolor{ioccshade}\textbf{Ka} &
\cellcolor{ioccshade}\textbf{S} &
\cellcolor{ioccshade}\textbf{AFS/LANS} &
\cellcolor{lightblue}\textbf{Ka} &
\cellcolor{lightblue}\textbf{S} &
\cellcolor{lightblue}\textbf{AFS/LANS} \\ \hline

\textbf{Number of links}
& 1 & 1 & 1
& 1 & 1 & 2 & 3
& 2 & 2 & 4
& 2 & 2 & 5 \\

\textbf{Fwd / Rtn link}
& R
& F+R
& F
& F+R
& F+R
& F
& F
& F+R
& F+R
& F
& F+R
& F+R
& F \\

\textbf{Service volume}
& \multicolumn{3}{c|}{SV1}
& \multicolumn{4}{c|}{SV1}
& \multicolumn{3}{>{\columncolor{ioccshade}[\tabcolsep][\tabcolsep]}c|}{SV2}
& \multicolumn{3}{c|}{SV3} \\

\textbf{Min coverage}
& {} & 70\% & {}
& 75\%
& 90\%
& 70\%
& 40\%
& 75\%
& 90\%
& 40\%\tnote{\dag}
& 75\%
& 95\%
& 99\% \\ \hline

\end{tabular}%
}
\begin{tablenotes}[flush]\footnotesize
\item[\dag] Minimum fraction of an Earth day with $\mathrm{GDOP}<6$ over SV2; GDOP threshold TBR (LCRNS.3.0130).
\end{tablenotes}
\end{threeparttable}
\end{table}

%==============================================================================
\section{IOC-C Requirements and Evaluation Metrics}
\label{sec:requirements}
%==============================================================================
The IOC-C requirements can be reduced to three constellation-geometry tests outlined in \cref{tab:iocc_eval_metrics}. R1 and R2 use thresholds stated directly by the SRD. R3 interprets the SRD initial EVA-support requirement as two usable-PNT windows per Earth day and evaluates those windows at $\mathrm{GDOP}<6$ because the SRD does not assign a GDOP threshold to EVA duration.

\begin{table}[H]
\centering
\caption{IOC-C constellation-geometry requirements evaluated in this study. R1 and R2 use thresholds stated directly by the SRD; R3 applies the usable-PNT interpretation described in the text.}
\label{tab:iocc_eval_metrics}
\small
\renewcommand{\arraystretch}{1.2}
\begin{tabular}{p{1.0cm}p{5.0cm}p{9cm}}
\toprule
\textbf{ID} & \textbf{SRD driver} & \textbf{Evaluation criterion} \\
\midrule
\textbf{R1} &
S-band forward/return coverage in SV2 (LCRNS.3.0110/3.0113) &
Each surface point has LOS to at least two relay spacecraft for at least $90\%$ of an Earth day. \\ \\

\textbf{R2} &
LANS coverage and GDOP (LCRNS.3.0120/3.0130) &
Each surface point has at least four geometrically diverse AFS/LANS links with $\mathrm{GDOP}<6$ for at least $40\%$ of an Earth day. The SRD marks the GDOP threshold as TBR. \\ \\

\textbf{R3} &
Initial EVA PNT support (LCRNS.3.0240) &
Each surface point supports at least two EVA-support windows per Earth day, each lasting at least $4$~h nominally plus $1$~h contingency. Scored here at $\mathrm{GDOP}<6$ using the usable-PNT, distributable-contingency interpretation described in the text. \\
\bottomrule
\end{tabular}
\end{table}

The three metrics are evaluated pointwise over the surface portion of SV2, defined here as all lunar longitudes at latitudes south of $-75^\circ$. A constellation passes only if the worst-performing SV2 grid point satisfies each metric. The analysis compares five- and six-satellite tulip-shaped orbit constellations against a recently published Lunar Data Network (LDN) reference~\cite{brack2025aas25758}. The baseline LDN case uses five satellites in three elliptical lunar frozen orbit (ELFO) planes, referred to as the ``5-satellite LDN'' or ``LDN ELFO''; a six-satellite, six-plane configuration was also considered to assess potential efficiency gains.

Two of these three tests hinge on a single geometric quantity: R2 is scored directly on $\textrm{GDOP} < 6$, and R3 counts EVA windows during which GDOP remains below the same threshold. GDOP maps range-measurement error into combined position and clock-bias error. For $m$ satellites at positions $\vect r_i$ observed from $\vect r_{\text{obs}}$, define $\hat{\vect u}_i = (\vect r_{\text{obs}} - \vect r_i)/\rho_i$, with $\rho_i = \|\vect r_{\text{obs}} - \vect r_i\|$. In a pseudorange global navigation satellite system (GNSS)-style framework~\cite{massatt1990geometric,misraenge2011gps,kaplanhegarty2017understanding}, the $i$-th row of the Jacobian $\matr H$ is $[\hat{\vect u}_i^T,\, 1]$, where the final column represents clock bias. The standard GDOP equation is $\mathrm{GDOP} \;=\; \sqrt{\tr\bigl[(\matr H^T\matr H)^{-1}\bigr]}$,  $\sigma_{\text{pos+clk}} \;=\; \mathrm{GDOP}\cdot\sigma_p$, 
which requires at least four visible satellites and full column rank of $\matr H$. This four-satellite, full-rank condition is the geometric origin of the SRD's requirement for four diverse AFS/LANS links in R2, and it is the condition each candidate constellation must sustain over SV2. \Cref{sec:results} evaluates GDOP pointwise over the SV2 grid for every architecture considered.

%==============================================================================
\section{Candidate Constellations}
\label{sec:constellation}
%==============================================================================
\subsection{Seven-petal tulip constellations}
Tulip-shaped orbits are periodic Earth-Moon circular restricted three-body problem (CR3BP) orbits that trace closed, pole-centered lobes in the Moon-fixed frame. These lobes, or \emph{petals}, remain fixed in longitude when the orbital period is commensurate with lunar rotation, causing the ground track to repeat over the same surface region each period. This work uses the southern \(6\!:\!5\) sidereal-resonant seven-petal member introduced and characterized in prior tulip-orbit studies~\cite{koblick2023tulip,koblick2025tulip,koblick2026station,zhang2026cja,zhang2026time}.

\begin{figure}[!htb]
\centering
\includegraphics[
    width=\linewidth,
    trim=1cm 1.75cm 1cm 1.8cm,
    clip
]{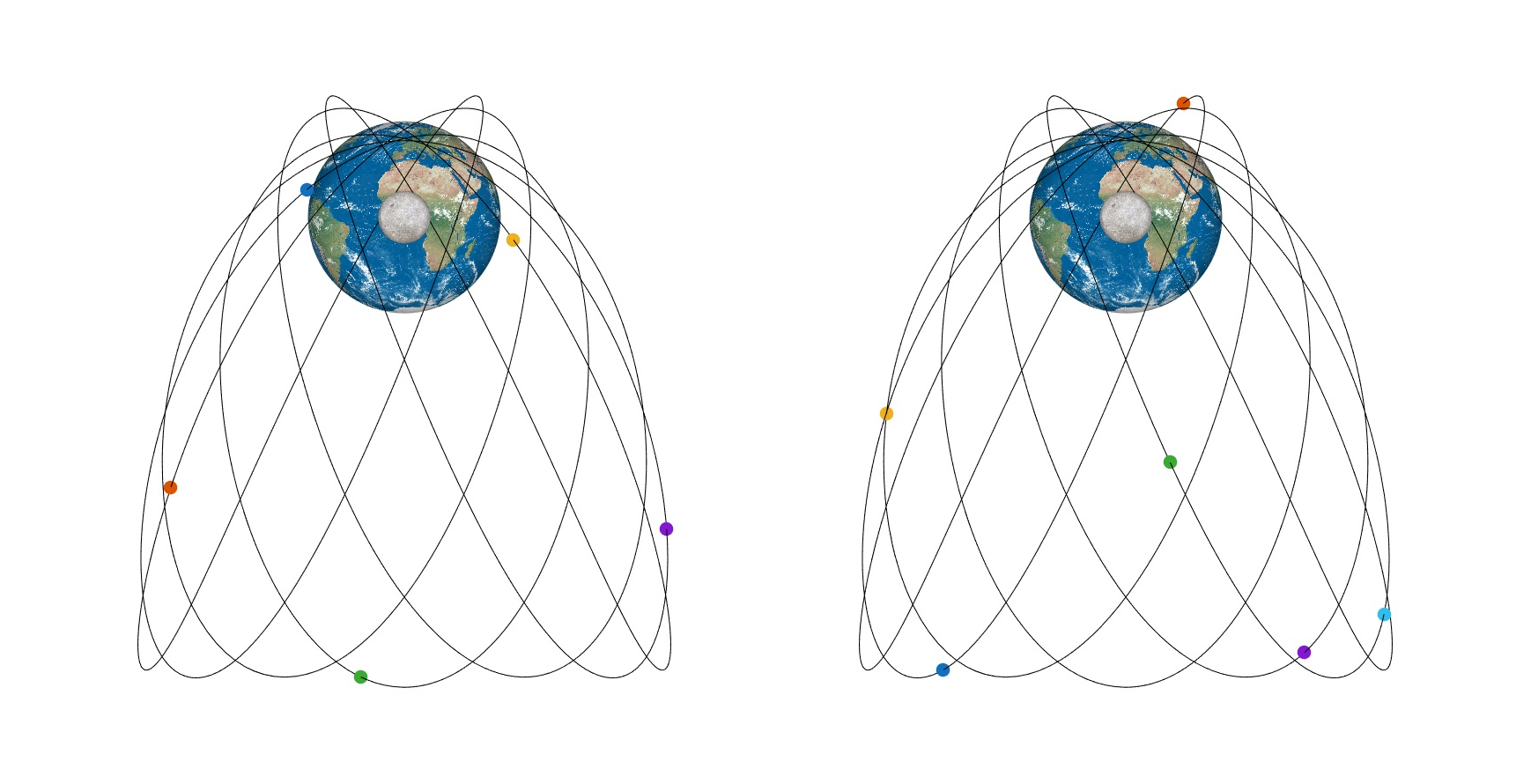}
\caption{Optimized tulip-shaped orbit constellations in the Earth-Moon barycentric rotating frame. \emph{Left:} the five-satellite configuration. \emph{Right:} the six-satellite configuration. Satellites occupy the \emph{same} shared sidereal-resonant tulip-shaped orbit, traced by the dark line over an orbital period ($\approx 23$~days). They differ in the number of satellites and phasing at $t=0$. The seven visible petals dip southward and concentrate sub-satellite passes over the lunar south-polar cap.}
\label{fig:constellation}
\end{figure}

All tulip constellation members occupy the same \(6\!:\!5\) sidereal-resonant three-body orbit, with nondimensional period $\tau_0 = 5\pi/3$ ($\approx 23$~days) and a seven-petal south-polar ground track~\cite{koblick2025tulip,koblick2026station}. Satellites differ only by along-orbit phase offsets $\{\Delta\tau_i\}$, expressed in nondimensional time~\cite{szebehely1967theory,koonlomarsdenross2011dynamical}. \Cref{fig:constellation} shows the five-satellite minimum-cost variant and the six-satellite recommended baseline.

\noindent The five- and six-satellite phasings are independent optima, obtained by maximizing the minimum EVA window, R3, over equally weighted observers at $-75^\circ$ latitude. This is accomplished with a genetic algorithm in the \textsc{pumpkyn} CR3BP toolbox~\cite{pumpkyn}. The resulting operating phasing vectors are
\begin{align}
\{\Delta\tau_i\}_{i=1}^{5} &= \{0.412,\,1.709,\,3.431,\,4.695,\,5.180\}, \nonumber\\
\{\Delta\tau_i\}_{i=1}^{6} &= \{0.778,\,1.146,\, 2.517,\, 2.893 ,\, 4.267,\, 4.632\} \quad \text{(dimensionless time)}.
\end{align}

%==============================================================================
\subsection{LDN-inspired ELFO reference case}
\label{sec:elfo}
\begin{figure}[H]
\centering
\vspace{-0.5em}
\includegraphics[
    width=\linewidth,
    trim=1cm 3cm 1cm 3.25cm,
    clip
]{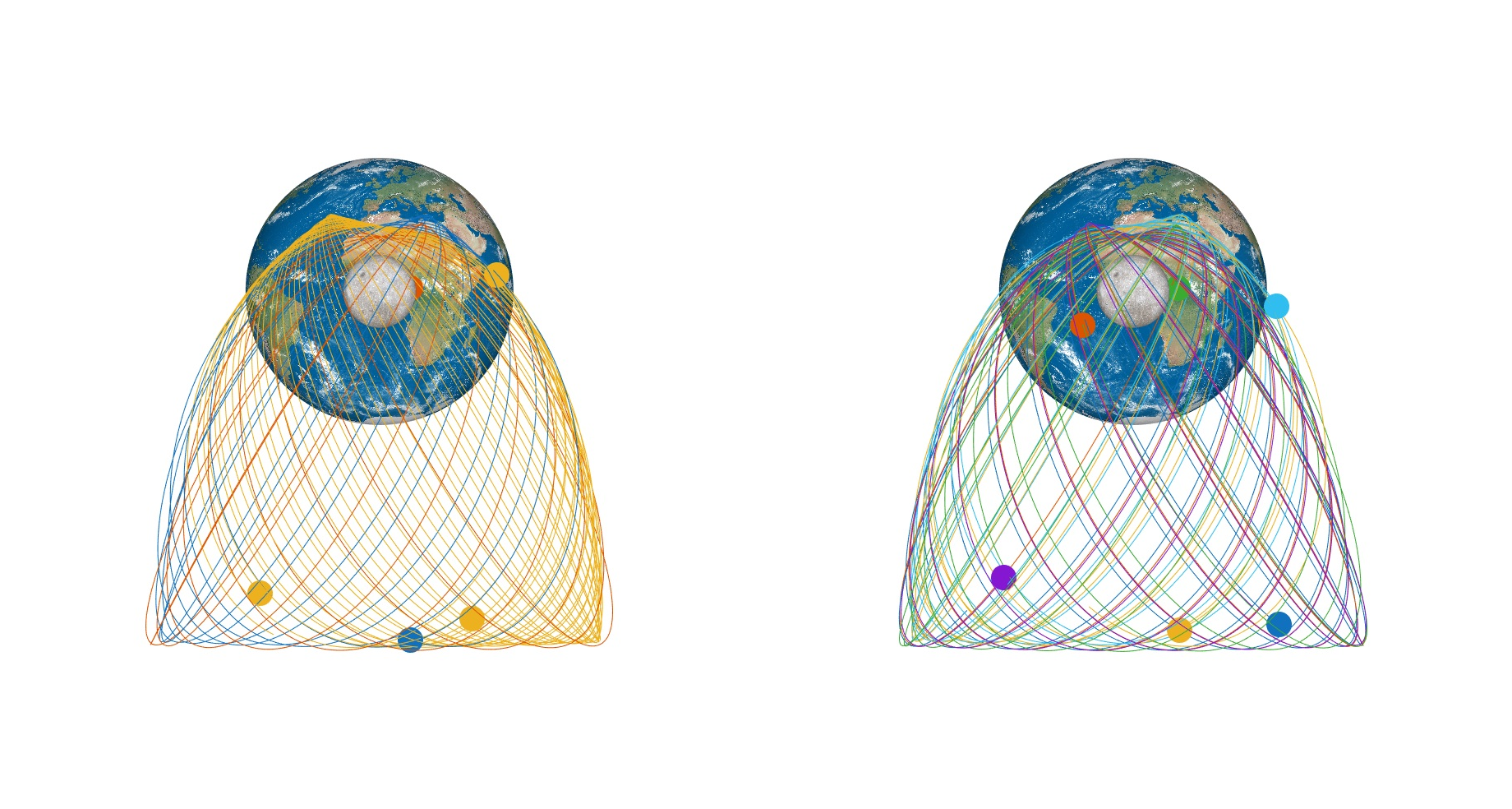}
\vspace{-0.75em}
\caption{The optimized ELFO Lunar Data Network in the Earth-Moon barycentric rotating frame. \emph{Left:} the five-satellite configuration in three planes, with apolune held over the southern hemisphere for south-polar dwell. \emph{Right:} the optimized six-satellite configuration in six separate planes.}
\label{fig:elfo}
\vspace{-0.5em}
\end{figure}

%==============================================================================
To benchmark the tulip constellations, we optimized two ELFO references: a five-satellite, three-plane case inspired by the published LDN design~\cite{brack2025aas25758}, and a six-satellite extension. Both cases use the same representative ELFO shape, with a semi-major axis of 12{,}000~$\mathrm{km}$, eccentricity of 0.69, inclination of $56.5^\circ$, and argument of perilune of $90^\circ$, placing apolune over the southern hemisphere. For the five-satellite case, the optimization varies the three RAAN values and five initial true anomalies while preserving the 1-1-3 satellite distribution across the three planes. For the six-satellite case, each spacecraft is assigned its own plane, and the optimization varies all six RAAN values and all six initial true anomalies (\cref{fig:elfo}).

\begin{table}[htbp]
\centering
\caption{Optimized ELFO constellation phasing variables for the 5-SV and 6-SV LDN configurations. Angles are in degrees. Within each plane, specified by the right ascension of the ascending node, $\Omega$, SVs are ordered by increasing initial true anomaly, $\nu_0$.}
\label{tab:elfo_phasing}
\footnotesize
\renewcommand{\arraystretch}{1.18}

\begin{threeparttable}
\resizebox{\linewidth}{!}{%
\begin{tabular}{
|>{\centering\arraybackslash}p{1.0cm}|
*{11}{>{\centering\arraybackslash}p{1cm}|}
}

\hline
&
\multicolumn{5}{c|}{\makecell{\textbf{ 5-SV LDN}, \textbf{3  planes}}} &
\multicolumn{6}{c|}{\makecell{\textbf{ 6-SV LDN}, \textbf{6  planes}}} \\
\hline

$\Omega$ [$^\circ$]
& 108.60
& 165.21
& \multicolumn{3}{c|}{269.50}
& 340.94
& 344.67
& 106.31
& 212.22
& 172.15
& 91.63 \\
\hline

$\nu_0$ [$^\circ$]
& 173.84
& 271.87
& 80.71
& 166.98
& 198.56
& 165.00
& 6.64
& 164.21
& 195.82
& 312.13
& 61.50 \\
\hline

\end{tabular}%
}
\end{threeparttable}
\end{table}

%==============================================================================
\section{Results}
\label{sec:results}
%==============================================================================

All architectures are evaluated over an SV2 grid spanning lunar latitudes from $-90^\circ$ to $-75^\circ$, all longitudes, and altitudes from 0 to 200~km. 
The time history is sampled at 5-minute time steps over a 23-day scenario (period of a \(6\!:\!5\) resonant three-body orbit), and all daily metrics are computed over sliding 24-hour windows.  Surface visibility uses a $0^\circ$ spherical-horizon elevation mask; terrain masking,  link-budget closure, antenna pointing, and multipath are not modeled. A satellite is counted as available for R1 when it has unobstructed line of sight to a user, and it is counted for R2/R3 when at least four visible satellites produce a full-rank GDOP solution with $\mathrm{GDOP}<6$.

The tulip phasing vectors are optimized with a genetic algorithm using a population size of 96, 40 stall generations, and a crossover fraction of 0.25. The objective maximizes the worst-case EVA usable-PNT metric over a stencil consisting of 25 surface points uniformly spaced in longitude at $-75^\circ$ latitude. The ELFO references use the same stencil and objective, but optimize RAAN and initial true anomaly while holding the ELFO orbital elements fixed. After optimization, all architectures are evaluated over the full SV2 grid using identical visibility, GDOP, and daily-window definitions. The evaluation grid contains six altitude layers from 0 to 200~km in 40~km increments. Each layer uses 64 observer points placed on latitude rings, with the number of longitude samples per ring scaled by $\cos(\mathrm{lat})$ to approximate equal-area spacing.

\subsection{Scorecard}
\Cref{tab:short_results} summarizes the pass/fail results and key architecture attributes. Both tulip constellations satisfy all three evaluated IOC-C geometry metrics over SV2. The six-satellite baseline provides wide margin, while the five-satellite variant passes as a minimum-cost case with less margin.  Beyond the required metrics, both tulip-shaped orbit constellations halve the total inter-satellite (sat--sat) occultation time relative to their ELFO counterparts and eliminate Earth occultation entirely, so any surface-visible spacecraft also has geometric access to Earth.

\begin{table}[htbp]
\centering
\caption{Architecture performance attributes for the five- and six-satellite tulip-shaped and ELFO~\cite{brack2025aas25758} configurations, measured against the LunaNet IOC-C requirements. All values are worst-SV2-point statistics; the occultation rows are diagnostic attributes with no SRD threshold.}
\label{tab:short_results}
\footnotesize
\renewcommand{\arraystretch}{1.18}

\begin{threeparttable}
\resizebox{\linewidth}{!}{%
\begin{tabular}{
@{}
>{\raggedright\arraybackslash}p{4.4cm}
>{\centering\arraybackslash}p{2.1cm}
>{\centering\arraybackslash}p{2.0cm}
>{\centering\arraybackslash}p{2.0cm}
>{\centering\arraybackslash}p{2.0cm}
>{\centering\arraybackslash}p{2.0cm}
@{}
}

\toprule
\textbf{Performance Metric} &
\makecell{\textbf{LunaNet}\\\textbf{IOC-C req.}} &
\makecell{\textbf{6-SV Tulip}\\\textbf{(baseline)}} &
\makecell{\textbf{6-SV LDN}\\\textbf{(ELFO)}} &
\makecell{\textbf{5-SV Tulip}\\\textbf{(min-cost)}} &
\makecell{\textbf{5-SV LDN}\\\textbf{(ELFO)}} \\
\midrule

R1: link availability
& $\geq 90\%$/day
& \textbf{100\%}
&\textbf{100\%}
&\textbf{100\%}
& \textbf{100\%} \\

R2: $\mathrm{GDOP}<6$
& $\geq 40\%$/day
& \textbf{75.35\%}
& \textbf{55.90\%}
& \textbf{43.61\%}
&\textbf{46.87\%} \\

R3: EVA continuity\tnote{a}
& $\geq 9$~h/day
&\textbf{18.00~h}
& \textbf{12.33~h}
& \textbf{9.92~h}
&\textbf{9.75~h} \\
\midrule

Max EVA gap time\tnote{b}
&  --
& 5.9~h
& 14.3~h
& 23.3~h
& 27.0~h \\

\midrule

Total Earth occultation\tnote{c}
& --
& 0~h
& 18.3~h
& 0~h
& 18.4~h \\

Total Sat--Sat occultation\tnote{d}
& --
& 81.5~h
& 173.5~h
& 77.1~h
& 147.8~h \\

\bottomrule

\end{tabular}%
}
\begin{tablenotes}[flushleft]\footnotesize
\item[a] Worst-SV2-point two-block budget, distributed-contingency reading at $\mathrm{GDOP}<6$; see Sec.~\ref{sec:requirements}.
\item[b] Derived metric: maximum time between $\mathrm{GDOP}<6$ access blocks lasting at least $5$~h; lower is better.
\item[c] Duration of time any satellite is occulted from Earth; lower is better.
\item[d] Duration of time any satellite LOS is occulted to another satellite; lower is better.
\end{tablenotes}
\end{threeparttable}
\end{table}

\subsection{Performance against IOC-C requirements (R1--R3)}
\label{sec:permetric}
\begin{figure}[H]
\centering
\captionsetup[subfigure]{
    justification=raggedright,
    singlelinecheck=false
}
\begin{subfigure}[b]{\linewidth}
\centering
\includegraphics[
    width=\linewidth,
    trim=0.25cm 0.5cm 0cm 0cm,
    clip
]{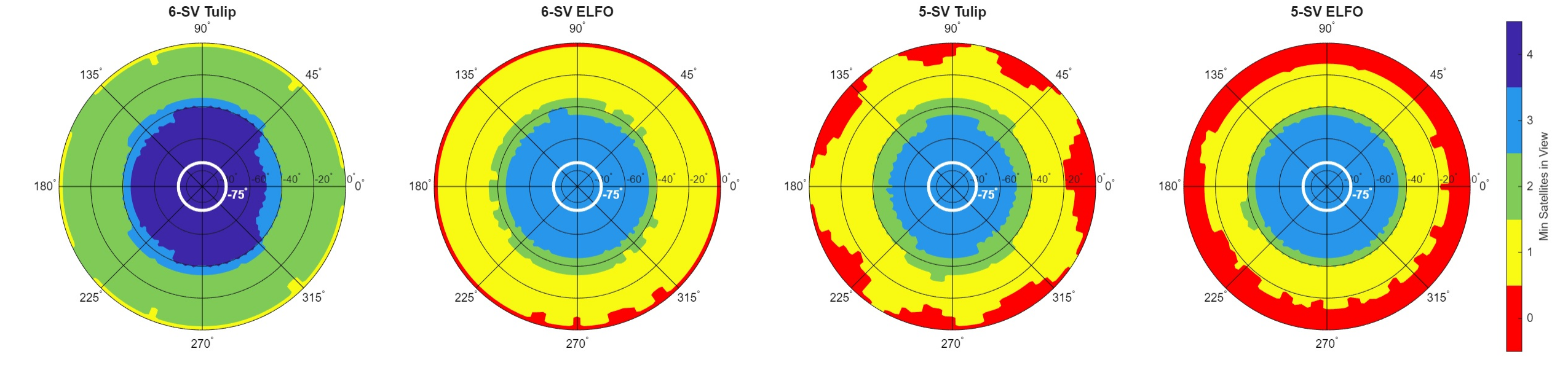}
\subcaption{R1: worst-case minimum number of satellites with LOS to a surface observer.}
\label{fig:los_hemisphere}
\end{subfigure}\\[0.9em]
\begin{subfigure}[b]{\linewidth}
\centering
\includegraphics[
    width=\linewidth,
    trim=0.25cm 0.5cm 0cm 0cm,
    clip
]{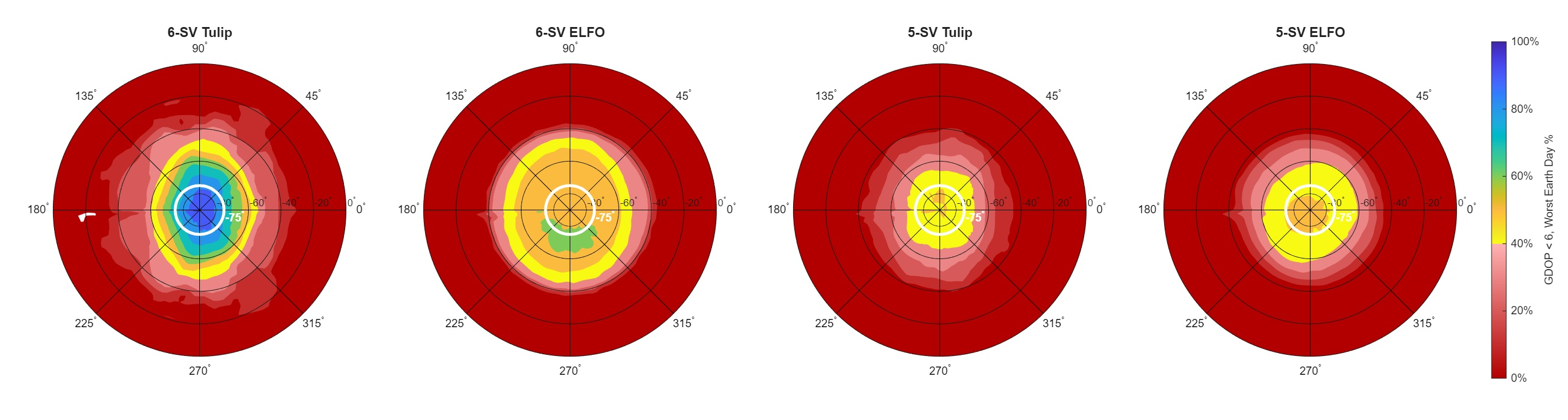}
\subcaption{R2: worst-day fraction of an Earth day with $\mathrm{GDOP}\!<\!6$ and $\geq\!4$ orbiters in view. Yellow satisfies the 40\% requirement; red violates it.}
\label{fig:r2_hemisphere}
\end{subfigure}\\[0.9em]
\begin{subfigure}[b]{\linewidth}
\centering
\includegraphics[
    width=\linewidth,
    trim=0.25cm 0.5cm 0cm 0cm,
    clip
]{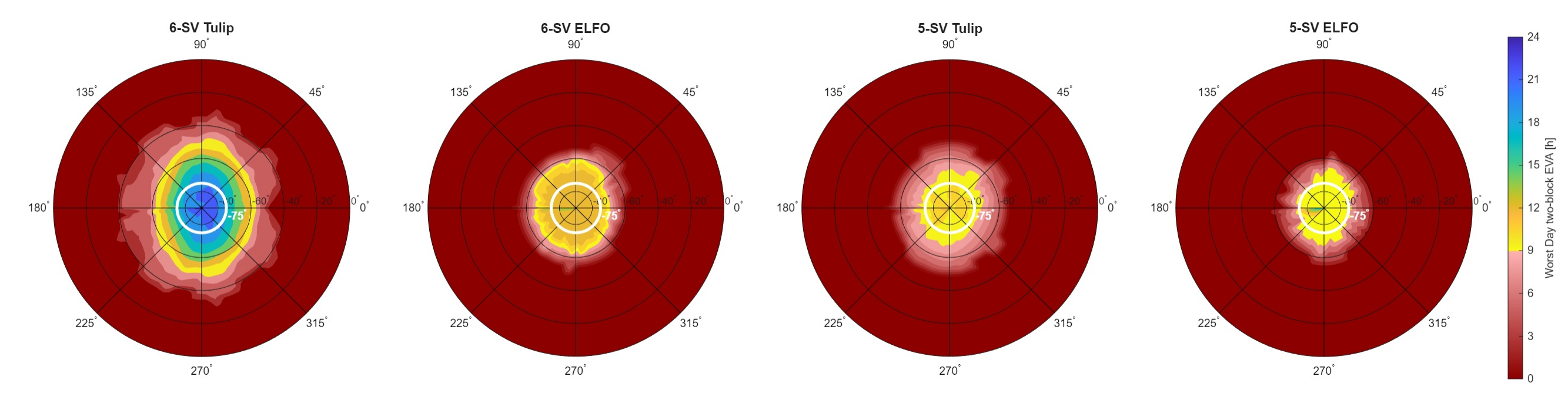}
\subcaption{R3: worst-day two-block EVA budget (two $\geq\!4$-h blocks totaling $\geq\!9$~h of $\mathrm{GDOP}\!<\!6$ geometry, distributed-contingency reading). Red fails the requirement.}
\label{fig:r3_hemisphere}
\end{subfigure}
\caption{IOC-C metric performance across the lunar southern hemisphere for the tulip-shaped and ELFO constellations. The white ring in each polar plot marks the SV2 boundary at $-75^\circ$ latitude. Maps are rendered at the lunar surface.}
\label{fig:iocc_maps}
\end{figure}

\Cref{fig:iocc_maps} maps all three IOC-C metrics across the lunar southern hemisphere for the four architectures. A constellation can meet an aggregate coverage fraction yet fail operationally if its SV2 access fragments into unusably short intervals; the contours therefore emphasize thresholded, worst-day behavior, showing where margin exists beyond the SV2 boundary.

\emph{R1 (link availability)} requires at least two satellites in line of sight for at least 90\% of each Earth day. All four architectures meet this requirement at every SV2 point, with 100\% two-satellite availability. The six-satellite tulip also maintains at least four visible satellites at all times and remains R1-compliant across the full southern hemisphere, while the five-satellite tulip and both ELFO cases fall below the 90\% threshold near $-40^\circ$ latitude.

\emph{R2 (LANS geometry)} requires $\mathrm{GDOP}<6$ and four satellites in view for $\ge40\%$ of an Earth day. All configurations pass across SV2 (\cref{fig:r2_hemisphere}). The worst-point daily availability is 75.35\% for the six-satellite tulip, 43.61\% for the five-satellite tulip, 55.90\% for the six-satellite LDN ELFO, and 46.87\% for the five-satellite LDN ELFO.

\emph{R3 (EVA usable-PNT windows)} is scored as two $\geq\!4$~h blocks totaling at least 9~h per day at $\mathrm{GDOP}<6$, served either by one continuous window or by two separate windows. All four configurations pass the surface portion within SV2. The worst-point two-block EVA budget is 18.00~h for the six-satellite tulip, 9.92~h for the five-satellite tulip, 12.33~h for the six-satellite LDN ELFO, and 9.75~h for the five-satellite LDN ELFO. \Cref{fig:r3_hemisphere} shows that the six-satellite tulip provides the largest usable-PNT region, while the five-satellite tulip and ELFO configurations pass over smaller caps.

\section{Operational \(\Delta V\) Screening}

This analysis addresses whether the tulip orbit is reachable from a practical launch condition, whether constellation phasing can be established without large inter-plane transfers, and how tulip-shaped orbit reconstitution compares with ELFO phasing maintenance. For both architectures, the screening is scoped to transfer feasibility, reconstitution, and orbit maintenance.

A representative low-thrust transfer was computed from an equatorial geostationary transfer orbit with perigee and apogee altitudes of \(350~\mathrm{km}\) and \(35{,}786~\mathrm{km}\), respectively, to the seven-petal southern \(6\!:\!5\)-resonant tulip-shaped orbit. The transfer was formulated in the CR3BP using an indirect minimum-time optimal-control approach~\cite{Zhang2015LowThrustRTBP}. For a \(15~\mathrm{kg}\) spacecraft with \(T_{\max}=25~\mathrm{mN}\) and \(I_{\mathrm{sp}}=2100~\mathrm{s}\), the transfer time is \(27.89~\mathrm{days}\), with an onboard electric-propulsion equivalent \(\Delta V\) of \(4.46~\mathrm{km/s}\) and \(2.92~\mathrm{kg}\) of propellant. Equivalently, the transfer requires an average acceleration capability of \(1.85~\mathrm{mm/s^2}\). Because the optimized trajectory is governed by the initial thrust-to-mass ratio and specific impulse, this result scales directly to heavier vehicles: a PNT/relay-class spacecraft of \(150~\mathrm{kg}\) equipped with \(250~\mathrm{mN}\) of thrust at the same \(I_{\mathrm{sp}}\) flies an identical \(27.89\)-day transfer with the same propellant mass fraction. The example therefore demonstrates compatibility with practical electric-propulsion performance across mission classes.

\begin{figure}[H]
\centering
\includegraphics[
    width=\linewidth,
    trim=0.9cm 0.25cm 2.0cm 0.40cm,
    clip
]{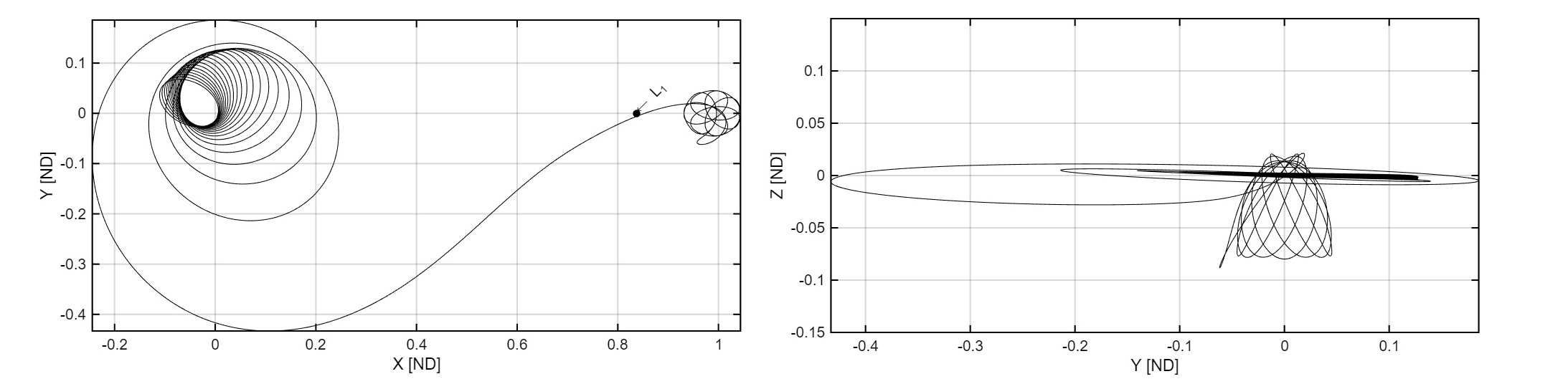}
\caption{Minimum-Time Low-thrust transfer from GTO to a seven-petal tulip-shaped orbit in the CR3BP. The left subplot is a polar view of the transfer, while the right subplot is looking from the Moon toward Earth.}
\label{fig:lowThrustXfer}
\end{figure}

Initial constellation phasing can be established by staggering the departure epoch of each spacecraft into the same shared tulip-shaped orbit. This avoids the need to deploy the constellation into multiple orbital planes. For post-failure reconstitution, a surviving spacecraft can temporarily transfer to a neighboring tulip-shaped orbit with a slightly different period, drift in phase, and return to the reference orbit. Shorter-period neighboring orbits advance the spacecraft phase; longer-period neighboring orbits lag behind it.

A theoretical lower bound on the impulsive transfer cost between neighboring tulip orbits can be estimated from their Jacobi constants as $\Delta V_{\min}
=
\left|
\sqrt{2U_{\max}-J_f}
-
\sqrt{2U_{\max}-J_i}
\right|,$ where \(U_{\max}\) is the maximum pseudo-potential over the two orbit trajectories, and \(J_i\) and \(J_f\) are the Jacobi constants of the initial and final orbits~\cite{whitley2018earth}. These values are useful as lower-bound diagnostics because they do not enforce endpoint geometry, maneuver timing, or cross-track velocity matching.

\begin{table}[H]
\centering
\caption{Minimum-energy lower-bound \(\Delta V\) estimates for transfers between the reference \(6\!:\!5\) tulip orbit and neighboring phasing orbits.}
\label{tab:min_energy_rephase_dv}
\small
\renewcommand{\arraystretch}{1.18}
\setlength{\tabcolsep}{2pt}

\begin{tabular*}{\linewidth}{@{\extracolsep{\fill}} l l l l c c c c c @{}}
\hline
\textbf{Phase} &
\textbf{Leg} &
\textbf{Initial \(p\!:\!q\)} &
\textbf{Target \(p\!:\!q\)} &
\(\boldsymbol{\Delta n\,[{}^\circ/\mathrm{day}]}\) &
\(\boldsymbol{J_i}\) &
\(\boldsymbol{J_f}\) &
\(\boldsymbol{U_{\max}}\) &
\(\boldsymbol{\Delta V_{\min}\,[\mathrm{m/s}]}\) \\
\hline

Advance &
Outbound &
\(6\!:\!5\) &
\(5\!:\!4\) &
+0.6463 &
3.1866 &
3.1828 &
\multirow{2}{*}{2.7795} &
\multirow{2}{*}{1.2524} \\

Advance &
Return &
\(5\!:\!4\)  &
\(6\!:\!5\) &
 &
3.1828 &
3.1866 &
 &
 \\

\hline

Lag &
Outbound &
\(6\!:\!5\) &
\(8\!:\!7\)  &
-0.7386 &
3.1866 &
3.1885 &
\multirow{2}{*}{2.2598} &
\multirow{2}{*}{0.8196} \\

Lag &
Return &
\(8\!:\!7\) &
\(6\!:\!5\) &
 &
3.1885 &
3.1866 &
 &
 \\

\hline

\multicolumn{8}{r}{\textbf{Total advance-phase lower bound}} &
\textbf{2.5048} \\

\multicolumn{8}{r}{\textbf{Total lag-phase lower bound}} &
\textbf{1.6392} \\
\hline
\end{tabular*}
\end{table}

These lower bounds suggest that phase reconstitution between nearby tulip-shaped orbits may be low cost. They are consistent with high-fidelity station-keeping analysis of the closely related sidereal-resonant tulip-shaped families, which reports mean annual maintenance \(\Delta V\) of approximately \(6\)--\(15~\mathrm{m/s}\) across all fourteen families examined in an ephemeris-based dynamical environment~\cite{koblick2026station}. Although the \(6\!:\!5\)-resonant orbit considered here belongs to a different resonance, those results indicate that tulip-shaped orbit maintenance is achievable at modest annual cost. For the ELFO references, the analogous operational burden is maintaining relative along-track spacing against secular drift. Recent navigation studies show that mean-anomaly drift degrades spacing and GDOP, but can be bounded with small in-plane semi-major-axis corrections costing centimeters per second per maneuver~\cite{ceresoli2025acta}. The COMPASS Lunar Network Satellite study similarly preserved frozen-orbit spacing over 12 years using a \(5~\mathrm{km}\) semi-major-axis offset, with \(50~\mathrm{m/s}\) reserved for phasing and \(50~\mathrm{m/s}\) for lunar-orbit operations margin~\cite{oleson2012compass}. We therefore reserve \(100~\mathrm{m/s}\) per ELFO spacecraft for combined phasing and station-keeping, consistent with the scope defined above.
%==============================================================================
\section{Discussion, Conclusions, and Future Analysis}
\label{sec:discussion}
%==============================================================================
This study evaluated seven-petal, \(6\!:\!5\)-resonant tulip-shaped orbit constellations for lunar south-pole PNT and direct-to-Earth relay against the IOC-C geometry requirements. Under the geometry-only, \(0^\circ\) elevation-mask assumptions of this analysis, both tulip architectures satisfy all three evaluated metrics across SV2, but with very different margin. The six-satellite constellation passes with room to spare, providing 75\% worst-point daily \(\mathrm{GDOP}<6\) availability and 18~h of daily EVA-usable PNT support; the five-satellite constellation passes with essentially none.  Six satellites is the baseline, and five is a minimum-cost option only where risk tolerance permits thin margins against modeling error, degraded vehicles, or added operational constraints.

The optimized ELFO configurations also satisfy the evaluated requirements, but the tulip architecture offers two structural advantages. First, every tulip spacecraft maintains continuous Earth line of sight, so any satellite visible from the surface is a geometric single-hop relay to Earth. Second, the constellation is a single shared orbit: spacecraft differ only in along-track
phase, simplifying deployment, rephasing, and post-failure reconstitution relative to multi-plane ELFO architectures.

Follow-on work proceeds in three steps. First, validate the dynamics: confirm maintenance and phasing reserves with high-fidelity force-model propagation and maneuver optimization, extending the sidereal-resonant station-keeping results of~\cite{koblick2026station} to the \(6\!:\!5\) family. Second, convert geometry to performance: extend the GDOP analysis to
a navigation filter with realistic clock and measurement errors, close the S-band, Ka-band, and AFS link budgets including SRD multipath margin, and characterize eclipse durations for power and thermal sizing. Third, mature the mission design: transfer, insertion, launch, and deployment analyses to quantify phasing requirements, extending coverage from IOC-C/SV2 toward
EOC/SV3 across additional tulip and pumpkin orbit families~\cite{zhang2026cja}.
%==============================================================================

\begingroup
\footnotesize
\singlespacing
\setlength{\parskip}{0pt}
\bibliographystyle{unsrt}
\bibliography{refs}
\endgroup
\end{document}